\documentclass[pra,aps,groupaddress,prd]{revtex4}
\usepackage{amsfonts}
\usepackage{epsfig,amsmath,graphicx,amssymb,overpic,bm}
\usepackage{color,tikz,xcolor,extarrows}
\usetikzlibrary{arrows}

\def\be{\begin{equation}}
\def\ee{\end{equation}}
\def\bee{\begin{eqnarray}}
\def\ene{\end{eqnarray}}
\def\bes{\begin{subequations}}
\def\ees{\end{subequations}}

\newcommand{\bR}{{\bf R}}
\newcommand{\mQ}{\mathcal{Q}}
\newcommand{\bQ}{{\bf Q}}
\newcommand{\bbR}{{\mathbb R}}

\newcommand{\bbQ}{{\mathbb Q}}
\newcommand{\bPT}{{\mathcal P}{\mathcal T}}
\newcommand{\bPC}{{\mathcal P}{\mathcal C}}
\newcommand{\bP}{{\mathcal P}}
\newcommand{\bT}{{\mathcal T}}
\def\v{\vspace{0.1in}}

\begin{document}

\title{A novel  hierarchy of two-family-parameter equations: Local, nonlocal, and  mixed-local-nonlocal vector nonlinear Schr\"odinger equations}
\author{Zhenya Yan$^{a,b}$}
\email{zyyan@mmrc.iss.ac.cn}
\affiliation{$^a$Key Lab of Mathematics Mechanization, Academy of Mathematics and Systems Science,  Chinese Academy of Sciences, Beijing 100190, China\\
$^b$School of Mathematical Sciences, University of Chinese Academy of Sciences, Beijing 100049, China}
%\date{\today}

\vspace{0.1in}
\begin{abstract}
{\bf Abstract.}\,\,We use two families of parameters $\{(\epsilon_{x_j}, \epsilon_{t_j})\,|\,\epsilon_{x_j,t_j}=\pm1,\, j=1,2,...,n\}$ to introduce a unified novel two-family-parameter system (simply called $\mQ^{(n)}_{\epsilon_{x_{\vec{n}}},\epsilon_{t_{\vec{n}}}}$ system), connecting integrable local, nonlocal,  novel mixed-local-nonlocal, and other nonlocal vector nonlinear Schr\"odinger (VNLS) equations.
The $\mQ^{(n)}_{\epsilon_{x_{\vec{n}}},\epsilon_{t_{\vec{n}}}}$ system with $(\epsilon_{x_j}, \epsilon_{t_j})=(\pm 1, 1),\, j=1,2,...,n$ is shown to possess Lax pairs and infinite number of conservation laws. Moreover, we also analyze the $\bPT$ symmetry of the Hamiltonians with self-induced potentials. The multi-linear forms and some symmetry reductions  are also studied. In fact, the used two families of parameters can also be extended to the general case $\{(\epsilon_{x_j}, \epsilon_{t_j})|\epsilon_{x_j}=e^{i\theta_{x_j}}, \epsilon_{t_j}=e^{i\theta_{t_j}},\,
\theta_{x_j}, \theta_{t_j}\in [0, 2\pi),\, j=1,2,...,n\}$ to generate more types of nonlinear equations.  The idea used in this paper can also be applied to other local nonlinear evolution equations such that novel integrable and non-integrable nonlocal and mixed-local-nonlocal systems can also be found.

\vspace{0.1in}

\end{abstract}

%\pacs{05.45.Yv, 02.30.Ik, 42.65.Tg}
\maketitle

\baselineskip=12pt

%%%%%%%%%%%%%%%%%%%%%%%%%%%%%%%%%%%%%%%%%%%%%%%%%%%%%%%%%%%%%%%%%

\vspace{-0.28in}

\section{Introduction}

Since a family of non-Hermitian parity-time ($\bPT$)-symmetric Hamiltonians $H=p^2+x^2(ix)^{\nu}$ with $\nu$ being a real constant was first shown by Bender and Boettcher~\cite{pt} to admit entirely real spectra, the $\bPT$-symmetric subject has been paid more and more attention (see Refs.~\cite{pt,ptr} and references therein). Here the parity reflection operator $\mathcal{P}:$ $x\rightarrow -x$ is linear whereas the time reflection operator $\mathcal{T}:$\,  $t\rightarrow -t$,\, $i\rightarrow -i$ is anti-linear~\cite{pt,ptr}.
 On  one hand, some linear and nonlinear equations with a variety of $\bPT$-symmetric potentials have been studied such as the Scarf-II potential, harmonic-Gaussian potential, rational potential~\cite{scar, Muss}. On the other hand, the $\bPT$-symmetric local extensions of nonlinear wave equations
have been explored  such as the KdV equation~\cite{p-kdv}, Burgers equation and short-pulse equation~\cite{y-burger}. Recently, a new integrable nonlocal NLS equation  was presented~\cite{am}, in which the self-included potential is  $\bPT$ symmetric for the fixed time. After that, we first introduced the new two-parameter $(\epsilon_x,\epsilon_t)$ local and nonlocal vector NLS equations ($\mQ^{(n)}_{\epsilon_x,\epsilon_t}$ system)~\cite{yanaml15} and its general form
($\mathcal{G}^{(n)}_{\epsilon_x,\epsilon_t}$ system)~\cite{yanaml16}. Moreover,  $\mQ^{(n)}_{\epsilon_x,\epsilon_t}$ and $\mathcal{G}^{(n)}_{\epsilon_x,\epsilon_t}$ systems were shown to be Lax integrable for $(\epsilon_x,\epsilon_t)=(1,1),\, (-1,1)$. Particularly, we found that integrable nonlocal single- and two-component NLS equations had novel
rational soliton-like structures~\cite{yannonlocal}. The stable nonlinear modes for the nonlocal NLS equation under the $\bPT$-symmetric potentials were also found~\cite{yanchaos-nonlocal}. After that, other types of nonlocal equations were also studied~\cite{nonlocal}.

To the best of our knowledge, all nonlinearities in the known nonlinear integrable systems are either  local or nonlocal.  A natural problem is whether there  exist some nonlinear integrable systems with  {\it both local and nonlocal nonlinearities}.
In this paper, we will introduce two families parameters $\{(\epsilon_{x_j},\epsilon_{t_j})|j=1,2,...,n\}$ to first present new nonlinear evolution hierarchy
$\mQ^{(n)}_{\epsilon_{x_{\vec{n}}},\epsilon_{t_{\vec{n}}}}$. In particularly, a hierarchy of novel one-family-parameter nonlinear integrable systems was presented based on a new symmetry reduction in the no-reduction Lax pair of the AKNS hierarchy~\cite{akns}. The new integrable hierarchy contains the local, nonlocal, and mixed-local-nonlocal systems. Moreover, we study their infinite number of conservation laws, and symmetry reductions. This letter can be regarded as the further development of the previous series of papers about the local and nonlocal vector NLS equations~\cite{yanaml15,yanaml16}. It should be pointed out that the idea of two families of parameters used in this paper can also be applied to find new local, nonlocal, and mixed-local-nonlocal systems of other wave equations such as the higher-order AKNS hierarchy including vector KdV equation, vector mKdV equations, and vector sine/sinh-Gordon equations, vector DS equations, vector short-pulse equations, vector KP equation, discrete systems, etc..

\section{A unified new two-family-parameter nonlinear evolution hierarchy}

We here introduce a hierarchy of unified two-family-parameter $\{(\epsilon_{x_j},\epsilon_{t_j})|j=1,2,...,n\}$ nonlinear evolution equations (shortly called $\mQ^{(n)}_{\epsilon_{x_{\vec{n}}},\epsilon_{t_{\vec{n}}}}$ system)
\bee  \label{vnlsp}
i\bQ_t(x,t)=-\bQ_{xx}(x,t)+2\bQ(x,t)\bQ^{\dag}(\epsilon_{x_{\vec{n}}}x,\epsilon_{t_{\vec{n}}}t)\Lambda\bQ(x,t),\quad  x, t\in \mathbb{R},\,
\ene
 where $\bQ(x,t)=(q_1(x,t), \, q_2(x,t),\,\cdots,\, q_n(x,t))^T$ is a complex-valued column vector function, the vector function $\bQ(\epsilon_{x_{\vec{n}}}x,\epsilon_{t_{\vec{n}}}t)$ is defined by $\bQ(\epsilon_{x_{\vec{n}}}x,\epsilon_{t_{\vec{n}}}t)=(q_1(\epsilon_{x_1}x,\, \epsilon_{t_1}t),\, q_2(\epsilon_{x_2}x,\, \epsilon_{t_2}t),\, \cdots, \,q_n(\epsilon_{x_n}x, \, \epsilon_{t_n}t))^T$ with  $\epsilon_{x_j,t_j}=\pm 1$ being symmetric parameters,\,  $\Lambda={\rm diag}(\sigma_1,\, \sigma_2,\, \cdots, \sigma_n)$ with $\sigma_j=\pm 1$ denoting the real-valued self-focusing ($-$) and defocusing ($+)$ nonlinear interactions for the component $q_j(x,t)$, $\dag$ stands for  the transpose conjugate. Here $\Lambda$ can also be replaced by the general Hermitian matrix ${\rm M}$ with  ${\rm M}^{\dag}={\rm M}=(M_{ij})_{n\times n},\, |{\rm M}|\not=0$.

 The $\mQ^{(n)}_{\epsilon_{x_{\vec{n}}},\epsilon_{t_{\vec{n}}}}$ nonlinear wave system (\ref{vnlsp}) contains many types of distinct evolution equations for two families of parameters $(\epsilon_{x_j}, \epsilon_{t_j})\in\{(1,1), (-1,1), (1, -1), (-1, -1)\},\,j=1,2,...,n$ such as the local~\cite{akns}, nonlocal~\cite{am,yanaml15,yanaml16}, and {\it novel} mixed-local-nonlocal vector NLS equations. The $\mQ^{(n)}(\epsilon_{x_j}=\epsilon_{t_j}=1)$ system is only local integrable vector NLS equations including the well-known Manakov system~\cite{vnls}. We find that the self-induced potentials $2\bQ^{\dag}(\epsilon_{x_{\vec{n}}}x,\epsilon_{t_{\vec{n}}}t)\Lambda\bQ(x,t)$ with $(\epsilon_{x_j}, \epsilon_{t_j})\in\{(-1,1), (1, -1), (-1, -1)\}$ must {\it not} be real-valued functions, and differ from the local case that $2\bQ^{\dag}(\epsilon_{x_{\vec{n}}}x,\epsilon_{t_{\vec{n}}}t)\Lambda\bQ(x,t)$ must be a real-valued function for $\epsilon_{x_j}=\epsilon_{t_j}=1,\, j=1, 2,...,n$. The quasi-power defined by $Q_{\epsilon_{x_{\vec{n}}}, \epsilon_{t_{\vec{n}}}}(t)=\int_{-\infty}^{+\infty}\bQ^{\!\dag}(\epsilon_{x_{\vec{n}}}x,\epsilon_{t_{\vec{n}}}t)\bQ(x,t)dx$ is conserved during evolution, however the total power of Eq.~(\ref{vnlsp}) defined by $P_{\epsilon_{x_{\vec{n}}}, \epsilon_{t_{\vec{n}}}}(t)=\int_{-\infty}^{+\infty}|\bQ(x,t)|^2dx$ is {\it not} conserved during evolution except that the power $P_{+1,+1}(t)$ is conserved during evolution since $dP_{\epsilon_{x_{\vec{n}}}, \epsilon_{t_{\vec{n}}}}(t)/(dt)=-2i \int_{-\infty}^{+\infty}\!\!dx|\bQ(x,t)|^2\![\bQ^{\!\dag}\!(\epsilon_{x_{\vec{n}}}x,\epsilon_{t_{\vec{n}}}t)\Lambda\bQ(x,t)\!\!-\!\!
\bQ^{\!\dag}(x,\!t)\Lambda\bQ(\epsilon_{x_{\vec{n}}}x, \epsilon_{t_{\vec{n}}}t)]$. 

\v {\bf Remark 1.} Following our previous idea~\cite{yanaml15}, we can also choose two families of parameters 
$(\epsilon_{x_j}, \epsilon_{t_j})\in\{(\epsilon_{x_j}, \epsilon_{t_j})|\epsilon_{x_j}=\pm 1, \pm i, \epsilon_{t_j}=\pm 1, \pm i,\, j=1,2,...,n\}$ to generate new nonlinear equations from system (\ref{vnlsp}). In general, we may also use the two families of parameters 
\bee \label{para}
(\epsilon_{x_j}, \epsilon_{t_j})\in\{(\epsilon_{x_j}, \epsilon_{t_j})|\epsilon_{x_j}=e^{i\theta_{x_j}}, \epsilon_{t_j}=e^{i\theta_{t_j}},\,
\theta_{x_j}, \theta_{t_j}\in [0, 2\pi),\, j=1,2,...,n\}
\ene
such that system (\ref{vnlsp}) or other local systems with the two families of parameters can generate more types of nonlinear wave equations.\\

We  analyze $\mQ^{(n)}_{\epsilon_{x_{\vec{n}}},\epsilon_{t_{\vec{n}}}}$ model with $n=2$ and $(\epsilon_{x_j}, \epsilon_{t_j})\in\{(1,1), (-1,1), (1, -1), (-1, -1)\},\, j=1,2$ in details:

  (i) For $n=1$, Eq.~(\ref{vnlsp}) yields the self-focusing $(\sigma_1=-1$) (defocusing $(\sigma_1=1$)) local ($\epsilon_{x_1}=\epsilon_{t_1}=1$)~\cite{am} or nonlocal ($(\epsilon_{x_1},\,\epsilon_{t_1})\in\{(-1,1),\,(1,-1),\, (-1,-1)\}$) NLS equation ($\mQ^{(1)}_{\epsilon_{x_{\vec{1}}},\epsilon_{t_{\vec{1}}}}$ model)~\cite{yanaml15}.

(ii) For $n=2$, the $\mQ^{(2)}_{\epsilon_{x_{\vec{2}}},\epsilon_{t_{\vec{2}}}}$ system given by Eq.~(\ref{vnlsp}) generates
\bee\label{nls2} \begin{array}{l}
 \mQ^{(2)}_{\epsilon_{x_{\vec{2}}},\epsilon_{t_{\vec{2}}}}:\, iq_{jt}(x,t)\!=\!-q_{jxx}(x,t)+2[\sigma_1q_1(x, t)q_1^{*}(\epsilon_{x_1} x,\,\epsilon_{t_1}t)
       +\sigma_2 q_2(x,t)q^{*}_2(\epsilon_{x_2} x,\, \epsilon_{t_2}t)]q_j(x,t), \,\, j=1,2,
  \end{array}\ene
where the star denotes the complex conjugate, which differs from our previous models~\cite{yanaml15,yanaml16} and contains two families of parameters $\{(\epsilon_{x_j},\, \epsilon_{t_j})|\epsilon_{x_j}=\pm 1,\, \epsilon_{t_j}=\pm 1, \, j=1,2\}$ to yield the known two-component models~\cite{yanaml15} and {\it new} nonlocal vector NLS equations. We discuss them as follows:
\begin{itemize}

\item {} For $\epsilon_{x_1}=\epsilon_{x_2}=\pm 1,\, \epsilon_{t_1}=\epsilon_{t_2}=\pm 1$ and $\sigma_1=\sigma_2=\pm 1$, system (\ref{nls2}) reduces to the known self-focusing or defocusing local Manokov systems~\cite{vnls}
   \bee\label{nls2-1} \begin{array}{l}
iq_{jt}(x,t)\!=\!-q_{jxx}(x,t)\pm 2[|q_1(x, t)|^2+ |q_2(x,t)|^2]q_j(x,t), \,\, j=1,2,
  \end{array}\ene
and known $q_1$-nonlocal-$q_2$-nonlocal systems~\cite{yanaml15,yanaml16}
   \bee\label{nls2-2} \begin{array}{l}
iq_{jt}(x,t)\!=\!-q_{jxx}(x,t)\pm 2[q_1(x, t)q_1^{*}(-x,\,t)+q_2(x, t)q_2^{*}(-x,\,t)]q_j(x,t), \,\, j=1,2, \v\\
iq_{jt}(x,t)\!=\!-q_{jxx}(x,t)\pm 2[q_1(x, t)q_1^{*}(x,\,-t)+ q_2(x, t)q_2^{*}(x,\,-t)]q_j(x,t), \,\, j=1,2, \v\\
iq_{jt}(x,t)\!=\!-q_{jxx}(x,t)\pm 2[q_1(x, t)q_1^{*}(-x,\,-t)+ q_2(x, t)q_2^{*}(-x,\,-t)]q_j(x,t), \,\, j=1,2,
  \end{array}\ene

\item {} For $\epsilon_{x_1}=\epsilon_{x_2}=\pm 1,\, \epsilon_{t_1}=\epsilon_{t_2}=\pm 1$, system (\ref{nls2}) reduces to the
known general local and nonlocal systems including both the same coefficients (see Eqs.(\ref{nls2-1})-(\ref{nls2-2})) and different coefficients of nonlinear terms~\cite{yanaml16}.

\item {} For the other cases, we find {\it novel mixed local-nonlocal} vector NLS equations, i.e., the $q_1$-nonlocal-$q_2$-local (or $q_1$-local-$q_2$-nonlocal) systems
\bee\label{nls2-3} \begin{array}{l}
iq_{jt}(x,t)\!=\!-q_{jxx}(x,t)+2[\sigma_1q_1(x, t)q_1^{*}(-x,\,t)+\sigma_2 |q_2(x,t)|^2]q_j(x,t), \,\, j=1,2, \v\\
iq_{jt}(x,t)\!=\!-q_{jxx}(x,t)+2[\sigma_1q_1(x, t)q_1^{*}(x,\, -t)+\sigma_2 |q_2(x,t)|^2]q_j(x,t), \,\, j=1,2, \v\\
iq_{jt}(x,t)\!=\!-q_{jxx}(x,t)+2[\sigma_1q_1(x, t)q_1^{*}(-x,\,-t)+\sigma_2 |q_2(x,t)|^2]q_j(x,t), \,\, j=1,2,
  \end{array}\ene
and {\it new} $q_1$-nonlocal-$q_2$-nonlocal systems
\bee\label{nls2-4} \begin{array}{l}
iq_{jt}(x,t)\!=\!-q_{jxx}(x,t)+2[\sigma_1q_1(x, t)q_1^{*}(-x,\,t)+\sigma_2 q_2(x, t)q_2^{*}(x,\,-t)]q_j(x,t), \,\, j=1,2, \v\\
iq_{jt}(x,t)\!=\!-q_{jxx}(x,t)+2[\sigma_1q_1(x, t)q_1^{*}(x,\, -t)+\sigma_2 q_2(x, t)q_2^{*}(-x,\,-t)]q_j(x,t), \,\, j=1,2, \v\\
iq_{jt}(x,t)\!=\!-q_{jxx}(x,t)+2[\sigma_1q_1(x, t)q_1^{*}(-x,\,t)+\sigma_2 q_2(x, t)q_2^{*}(-x,\,-t)]q_j(x,t), \,\, j=1,2,
  \end{array}\ene
which differ from the known $q_1$-nonlocal-$q_2$-nonlocal system (\ref{nls2-2})~\cite{yanaml15,yanaml16}.
 
 \v Notice that if we choose the general  two families of parameters given by Eq.~(\ref{para}), then we obtain more types of two-component nonlinear equations.
\end{itemize}
In fact, under the roles of two families of parameters $(\epsilon_{x_j}, \epsilon_{t_j}),\, j=1,2,...,n$, Eq.~(\ref{vnlsp}) with $n>1$  can generate more and more types of nonlinear wave equations as the number $n$ increases, some of which are {\it new} mixed-$m$-local-$(n-m)$-nonlocal and nonlocal vector NLS equations, where
$1\leq m<n$.

\section{Lax pair and infinite number of conservation laws}

Consider the linear iso-spectral system~\cite{yanaml15,yanaml16,vnlsb}
\vspace{-0.03in}
\bee\label{lax} \begin{array}{l}
\Psi_x+i\lambda \Sigma_3\Psi=U\Psi, \v\\
\Psi_t+2i\lambda^2\Sigma_3\Psi=[2\lambda U-i(U_x+U^2)\Sigma_3]\Psi,
\end{array}
\ene
where $\Psi=(\psi_1(x,t),\,\psi_2(x,t),...,\psi_{n+1}(x,t))^T$ is a $(n+1)\times 1$ column complex-valued eigenvector function, $\lambda\in \mathbb{C}$ is an iso-spectral parameter, the generalized Pauli matrix $\Sigma_3$  and potential matrix $U$ are defined by
\bee \nonumber
 \Sigma_3=\left(\begin{matrix}  I_{n\times n}  &  0 \\ 0 & -1 \end{matrix} \right), \quad
 U=U(x,t)= \left(\begin{matrix}  0_{n\times n} &  \bQ(x,t) \\  \bR(x,t) & 0 \end{matrix} \right), \,\,
\ene
where $I_{n\times n}$ and  $ 0_{n\times n}$ are $n\times n$ unity and zero matrixes, respectively, and $\bR(x,t)=(r_1(x,t), r_2(x,t),\cdots, r_n(x,t))$ is a complex-valued row vector function.  The compatibility condition of Eq.~(\ref{lax}), $\Psi_{xt}=\Psi_{tx}$, i.e., zero-curvature equation, leads to the general $2n$-component coupled nonlinear wave system
\bee \label{vnls}\begin{array}{l}
i\bQ_t(x,t)=-\bQ_{xx}(x,t)+2 \bQ(x,t)\bR(x,t)\bQ(x,t),  \v\\
-i\bR_t(x,t)=-\bR_{xx}(x,t)+2 \bR(x,t)\bQ(x,t)\bR(x,t).
\end{array}
\ene

In the following we consider some symmetry reductions of system (\ref{vnls}):

\begin{itemize}

\item {} {\it Symmetry reduction-I}: $\bR(x,t)=0$,  system (\ref{vnls}) becomes the linear Schr\"odinger equations without potentials $i\bQ_t(x,t)=-\bQ_{xx}(x,t)$.

\item {} {\it Symmetry reduction-II}: $\bR(x,t)=\sigma\bQ^{\dag}(x,t)$ with $\sigma=\pm 1$ which make system (\ref{vnls}) reduce to the known vector NLS equations~\cite{vnlsb}
\vspace{-0.05in}
\bee\label{nls-1}
i\bQ_t(x,t)=-\bQ_{xx}(x,t)+2\sigma \bQ(x,t)\bQ^{\dag}(x,t)\bQ(x,t), \quad
\ene
 which corresponds to Eq.~(\ref{vnlsp}) with $(\epsilon_{x_j}, \epsilon_{t_j})=(1,1)$ and $\Lambda=\sigma I_{n\times n}$ containing the self-focusing ($\sigma=-1)$ and defocusing ($\sigma=1)$ single-component NLS equation ($n=1$) and two-component Manakov system ($n=2$)~\cite{vnls}.

\item {} {\it Symmetry reduction-III}: $\bR(x,t)=\bQ^{\dag}(x,t)\Lambda$, which make system (\ref{vnls}) reduce to the vector NLS equations~\cite{vnlsb}
\vspace{-0.05in} \bee\label{nls-2}
i\bQ_t(x,t)=-\bQ_{xx}(x,t)+2\bQ(x,t)\bQ^{\dag}(x,t)\Lambda\bQ(x,t), \quad
\ene
For $\Lambda=I_{n\times n}$ or $\Lambda=-I_{n\times n}$, system (\ref{nls-2}) reduces to system (\ref{nls-1}), otherwise, system (\ref{nls-2}) is a mixed self-focusing and defocusing system.

\item{} {\it Symmetry reduction-IV}: Introduce a new one-family-parameter $\{\epsilon_{x_j}=\pm 1,\, j=1,2,...,n\}$ symmetry reduction
\bee\label{con}
 \bR(x,t)=\bQ^{\dag}(\epsilon_{x_{\vec{n}}}x, t)\Lambda,  \quad \epsilon_{x_j}=\pm 1,\quad \epsilon_{t_j}=1,\quad j=1,2,...,n,
\ene
which make system (\ref{vnls}) generate the above-introduced new local-nonlocal vector equations (\ref{vnlsp}) with $\epsilon_{x_j}=\pm 1$ and
$\epsilon_{t_j}=1$, that is, Eq.~(\ref{vnlsp}) with $\epsilon_{x_j}=\pm 1$ and $\epsilon_{t_j}=1$ admits the Lax pair given by Eq.~(\ref{lax}) with the new symmetric constraint (\ref{con}).

\item{} {\it Symmetry reduction-V}: we introduce a general one-family-parameter $\{\epsilon_{x_j}| \epsilon_{x_j}=\pm 1,\, j=1,2,...,n\}$ symmetry reduction with a Hermite matrix ${\rm M}$, $\bR(x,t)=\bQ^{\dag}(\epsilon_{x_{\vec{n}}}x, t){\rm M},  \quad  {\rm M}^{\dag}={\rm M}=(M_{ij})_{n\times n},\quad |{\rm M}|\not=0,\quad \epsilon_{x_j}=\pm 1,\quad \epsilon_{t_j}=1,\quad j=1,2,...,n$, which generates the general system
\bee  \label{vnlspm}
{\cal G}^{(n)}(\epsilon_{x_{\vec{n}}}, 1): \quad i\bQ_t(x,t)=-\bQ_{xx}(x,t)+2\bQ(x,t)\bQ^{\dag}(\epsilon_{x_{\vec{n}}}x, t){\rm M}\bQ(x,t),\quad  x, t\in \mathbb{R},\,
\ene
In particular, for the special case ${\rm M}=\Lambda$, system (\ref{vnlspm}) reduces to system (\ref{vnlsp}).
\end{itemize}

 For $n=2$, we known that Eq.~(\ref{nls2-1}) for the Manakov system, the first one for the nonlocal-nonlocal system of Eq.~(\ref{nls2-2}), and the first one for the nonlocal-local system of Eq.~(\ref{nls2-3}) are Lax integrable.
For $n=3$, except for the known integrable local~\cite{vnlsb} and nonlocal~\cite{yanaml15,yanaml16} systems,  we here find two families of
{\it novel integrable} mixed-local-nonlocal three-component NLS equations: the {\it mixed-$q_1$-nonlocal-$q_2$-local-$q_3$-local} system
\bee\label{nls3} \begin{array}{l}
iq_{jt}(x,t)\!=\!-q_{jxx}(x,t)+2\left[\sigma_1q_1(x, t)q_1^{*}(-x,\,t)+\sigma_2 |q_2(x,t)|^2+\sigma_3 |q_3(x,t)|^2\right]q_j(x,t), \,\,\, j=1,2,3,
  \end{array}\ene
and  the {\it mixed-$q_1$-nonlocal-$q_2$-nonlocal-$q_3$-local} system
\bee\label{nls3} \begin{array}{l}
iq_{jt}(x,t)\!=\!-q_{jxx}(x,t)+2\left[\sigma_1q_1(x, t)q_1^{*}(-x,\, t)+\sigma_2q_2(x, t)q_2^{*}(-x,\, t)+\sigma_3 |q_2(x,t)|^2\right]q_j(x,t),\, \,\, j=1,2,3,
  \end{array}\ene

Notice that for the other cases that at least one $\epsilon_{t_s}=-1,\, s\in \{1,2,...,n\}$,  the Lax pair of Eq.~(\ref{vnlsp}) is unknown yet such as  the last two ones for the nonlocal-nonlocal systems of Eq.~(\ref{nls2-2}), the last two ones for the nonlocal-local systems of Eq.~(\ref{nls2-3}), and Eq.~(\ref{nls2-4}) for the nonlocal-nonlocal systems.

In the following we only consider system (\ref{vnlsp}) with the Lax integrable case for $\epsilon_{x_j}=\pm 1,\, \epsilon_{t_j}=1,\, j=1,2,...,n$.
We introduce $n$ new complex functions~\cite{wadati} $\omega_j(x,t)=\psi_j(x, t)/\psi_{n+1}(x, t)\, (j=1,2,...,n)$ in terms of $n+1$ eigenfunctions $\psi_j(x,t)$ of Eq.~(\ref{lax}) such that we find that $\omega_j(x,t)$'s satisfy the $n$-component Riccati equations
\vspace{-0.03in}\bee
 \omega_{j,x}(x,t)\!=q_j(x,t)-\Big[2i\lambda +\sum_{k=1}^n\sigma_kq_k^{*}(\epsilon_{x_k}x,t)\omega_k(x,t)\Big]\omega_j(x,t),\quad j=1,2,...,n,
\label{ri}\ene

To solve Eq.~(\ref{ri}) we here consider their asymptotic (in $\lambda$) solutions and assume
their candidate solutions as power series expansions in parameter $2i\lambda$, with unknown functions of space and time as coefficients
 in the form
 \bee
 \label{omega}
 \begin{array}{l}
 \displaystyle\omega_j(x,t)=\sum_{s=0}^{\infty}\frac{\omega_{j}^{(s)}(x,t)}{(2i\lambda)^{s+1}}=\frac{\omega_{j}^{(0)}(x,t)}{2i\lambda}
 +\frac{\omega_{j}^{(1)}(x,t)}{(2i\lambda)^{2}}+\cdots,
 \end{array}
  \ene
 where $\omega_{j}^{(s)}(x,t)$'s are unknown functions to be determined.  Substituting  this assume into Eq.~(\ref{ri}) and comparing coefficients of terms $(2i\lambda)^{s}\, (s=0,1,2,...)$ to find
\bee\begin{array}{l}
 \omega_{j}^{(0)}(x,t)=q_j(x,t), \,\,\, \omega_{j}^{(1)}(x,t)=-q_{j,x}(x,t),\vspace{0.05in} \\
  \omega_{j}^{(s+1)}(x,t)\!=\displaystyle\sum_{i=1}^n\sigma_iq_i^{*}(\epsilon_{x_i}x,t)\!\!\sum_{k=1}^{s-1}\omega_{j}^{(k)}(x,t)\omega_{i}^{(s-k)}(x,t)
 -\omega_{j,x}^{(s)}(x,t), \,\, (s=2,3,...)
\end{array}
\ene

It follows from Eq.~(\ref{lax}) with condition (\ref{con}) that $
 (\ln |\psi_{n+1}|)_x\!=\!i\lambda+F(x,t),\,  (\ln |\psi_{n+1}|)_t\!=\!2i\lambda^2+ G(x,t)$,
where $F(x,t)=\sum_{k=1}^n \sigma_kq_k^{*}(\epsilon_{x_k}x, t)\omega_k(x,t)$, and $G(x,t)=\sum_{k=1}^n\sigma_k[2\lambda q_k^{*}(\epsilon_{x_k}x,t)-i\epsilon_{x_k}q_{kx}^{*}(\epsilon_{x_k}x,t)]\omega_k(x,t)
   +i\sum_{j=1}^n\sigma_jq_j(x,t)q_j^{*}(\epsilon_{x_j}x, t)$.
The compatibility condition, $(\ln |\psi_{n+1}|)_{xt}=(\ln |\psi_{n+1}|)_{tx}$, yields
\bee\label{conver} F_t(x,t)=G_x(x,t),
\ene

Substituting Eq.~(\ref{omega}) into the conversed Eq.~(\ref{conver}) and comparing the coefficients of same terms $\lambda^j$ yields the infinite number of conservation laws. For example, the first conservation laws are given by
\bee\begin{array}{l} \nonumber
\partial_t\sum_{k=1}^n\sigma_{k}q_k(x,t)q_k^{*}(\epsilon_{x_k}x, t)=i\partial_x\sum_{k=1}^n\sigma_{k}[q_{kx}(x,t)q_{k}^{*}(\epsilon_{x_k}x, t)
 -\epsilon_{x_k} q_{k}(x,t)q_{kx}^{*}(\epsilon_{x_k}x, t)], \v\\
\partial_t\sum_{k=1}^n\sigma_{k}q_k(x,t)q_{kx}^{*}(\epsilon_{x_k}x, t)=i\partial_x\sum_{k=1}^n \sigma_{k}[q_{kx}(x,t)q_{kx}^{*}(\epsilon_{x_k}x, t)
  \!\!-\!\!\epsilon_{x_k} q_{k}(x,t)q_{kxx}^{*}(\epsilon_{x_k}x, t)\!\!-\!\!\epsilon_{x_k} q_{k}^2(x,t)q_{k}^{*2}(\epsilon_{x_k}x, t)],
\end{array}
 \ene
where $q_{kx}^{*}(\epsilon_{x_k}x, t)=q_{k\xi}^{*}(\xi, t)|_{\xi=\epsilon_{x_k}x}$.

Therefore, we know that $\mQ^{(n)}_{\epsilon_{x_{\vec{n}}},\epsilon_{t_{\vec{n}}}}$ system (\ref{vnlsp}) is an integrable system for the parameter choices $(\epsilon_{x_j}, \epsilon_{t_j})=(\pm 1, 1)$. Notice that the inverse scattering method~\cite{vnlsb} can also be extended to solve Eq.~(\ref{vnlsp}) by means of its Lax pair (\ref{lax}) with (\ref{con}), which will be presented in another literature.

We now consider another change of Lax pair (\ref{lax}). If we make the transformation $t\to i\tau$, then Lax pair (\ref{lax}) becomes the linear iso-spectral system
\vspace{-0.03in}
\bee\label{laxg} \begin{array}{l}
\Phi_x+i\lambda \Sigma_3\Phi=W\Phi, \v\\
\Phi_{\tau}-2\lambda^2\Sigma_3\Phi=[2i\lambda W+(W_x+W^2)\Sigma_3]\Phi,
\end{array}
\ene
with $
 W=W(x,\tau)= \left(\begin{matrix}  0_{n\times n} &  \bbQ(x,\tau) \\  \bbR(x,\tau) & 0 \end{matrix} \right)$,
 $\bbQ(x,\tau)=(q_1(x,\tau), q_2(x,\tau),..., q_n(x,\tau))^T$, and $\bbR(x,\tau)=(r_1(x,\tau), r_2(x,\tau),..., r_n(x,\tau))$
such that the compatibility condition of Eq.~(\ref{laxg}), $\Phi_{x\tau}=\Phi_{\tau x}$, i.e., zero-curvature equation, leads to the general $2n$-component system
\bee \label{vnlsg}\begin{array}{l}
\bbQ_{\tau}(x,\tau)=-\bbQ_{xx}(x,\tau)+2 \bbQ(x,\tau)\bbR(x,\tau)\bbQ(x,\tau),  \v\\
-\bbR_t(x,\tau)=-\bbR_{xx}(x,\tau)+2 \bbR(x,\tau)\bbQ(x,\tau)\bbR(x,\tau).
\end{array}
\ene

\begin{itemize}

\item {} For $\bbR(x,\tau)=0$, system (\ref{vnlsg}) becomes the  heat equations $\bbQ_{\tau}(x,\tau)=-\bbQ_{xx}(x,\tau)$;

\item {} For $\bbR(x,\tau)=\bbQ^T(\epsilon_{x_{\vec{n}}}x,-\tau){\mathbb M}$ with ${\mathbb M}^T={\mathbb M}$ and $|{\mathbb M}|\not=0$, then  system (\ref{vnlsg}) is a hierarchy of integrable nonlocal or local-nonlocal real vector  NLS equations
\bee\label{nis}
\bbQ_{\tau}(x,\tau)=-\bbQ_{xx}(x,\tau)+2 \bbQ(x,\tau)\bbQ^T(\epsilon_{x_{\vec{n}}}x, -\tau){\mathbb M}\bbQ(x,\tau),
\ene
containing the two-component  case for $n=2$ and ${\mathbb M}=(m_{ij})_{2\times 2}$ as
\bee
\begin{array}{rl}
 q_{jt}(x,\tau)=& -q_{jxx}(x,\tau)+2\Big[(m_{11}q_1(\epsilon_{x_1}x, -\tau)+m_{12}q_2(\epsilon_{{x_2}}x, -\tau)q_1(x,\tau) \\
&\qquad +(m_{12}q_1(\epsilon_{x_1}x, -\tau)+m_{22}q_2(\epsilon_{{x_2}}x, -\tau))q_2(x,\tau)\Big]q_j(x,\tau), \quad j=1,2,\, \epsilon_{x_k}=\pm 1,\, k=1,2,
 %[(m_{12}q_1(\epsilon_{x_1}x, -\tau)+m_{22}q_2(\epsilon_{x_2}x, -\tau)]q_2(x,\tau)]]q_j(x,t),\quad j=1,2,
\end{array}
\ene
\end{itemize}
Similarly, we also find the infinite number of conservation laws of nonlocal integrable system (\ref{nis}).

In fact, we can also introduce  the general system (shortly called ${\mathbb Q}^{(n)}_{\epsilon_{x_{\vec{n}}},\epsilon_{\tau_{\vec{n}}}}$ system)
\bee
\bbQ_{\tau}(x,\tau)=-\bbQ_{xx}(x,\tau)+2 \bbQ(x,\tau)\bbQ^T(\epsilon_{x_{\vec{n}}}x,\epsilon_{\tau_{\vec{n}}}\tau){\mathbb M}\bbQ(x,\tau),
\ene
where $\bbQ(\epsilon_{x_{\vec{n}}}x,\epsilon_{\tau_{\vec{n}}}\tau)=(q_1(\epsilon_{x_1}x,\epsilon_{\tau_1}\tau), \, q_2(\epsilon_{x_2}x,\epsilon_{\tau_2}\tau), ...,
q_n(\epsilon_{x_n}x,\epsilon_{\tau_n}\tau))$ with $\epsilon_{x_j}$ and $\epsilon_{\tau_j}=\pm 1,\, j=1,2,...,n$ being defined by Eq.~(\ref{para}).

\v {\bf Remark 2.}  The idea of two families of parameters used in this paper can also be applied to find new local, nonlocal, and mixed-local-nonlocal systems of other wave equations such as the higher-order AKNS hierarchy including vector KdV equation, vector mKdV equations, and vector sine/sinh-Gordon equations, vector DS equations, vector short-pulse equations, vector KP equation, discrete systems, etc..

\section{The Hamiltonian with the self-included potentials and $\bPT$ symmetry}

In the following we investigate the $\bPT$ symmetrility of the solutions of  $\mQ^{(n)}_{\epsilon_{x_{\vec{n}}}, \epsilon_{t_{\vec{n}}}}$ system (\ref{vnlsp}).  For system (\ref{vnlsp}), we act the $\bPT$-symmetric operator on its both sides to yield
 \bee
 iq^{*}_{jt}(-x,-t)=-q^{*}_{jxx}(-x,-t)+2q_j^*(-x, -t)\sum_{k=1}^n \sigma_k q_k^{*}(-x,-t)q_k(\epsilon_{x_k}(-x), \epsilon_{t_k}(-t)),\quad j=1,2,...,n,
\ene
which for $q^{*}_{j}(-x,-t)$  are equivalent to Eq.~(\ref{vnlsp}) for $q_j(x,t)$, that is, if $q_j(x,t)$'s are solutions of Eq.~(\ref{vnlsp}), then
 their $\bPT$ relations $q_j^{*}(-x,-t)=\bPT q_j(x, t)$ also satisfy Eq.~(\ref{vnlsp}).  This implies that Eq.~(\ref{vnlsp}) is $\bPT$ symmetric for any parameter choices $(\epsilon_{x_j}, \epsilon_{t_j})$.

Eq.~(\ref{vnlsp}) can be rewritten as $i\bQ_t(x,t)=\hat{H}_n(\epsilon_{x_{\vec{n}}} x, \epsilon_{t_{\vec{n}}}t)\bQ(x,t)$, where the Hamiltonian operator $\hat{H}_n$ with self-induced potential is of the form
\bee \label{hn}
  \hat{H}_n(\epsilon_{x_{\vec{n}}} x, \epsilon_{t_{\vec{n}}}t)=-\partial_x^2+\sum_{k=1}^nV_k(x,t)=-\partial_x^2+2\sum_{k=1}^n\sigma_k q_k^*(\epsilon_{x_k}x,\epsilon_{t_k}t)q_k(x,t),
\ene
Now we consider the $\bPT$ symmetribility of the very component potential $V_k(x,t)$ in the Hamiltonian operators $\hat{H}_n$ (see Table~\ref{hnpt}), where the conjugate operator ${\cal C}$ is defined by ${\cal C}: i\to -i$. Notice that these terms including `No' in Table~\ref{hnpt} do not imply that those symmetries must not hold. They may admit some symmetries for some special potentials $g_k(x,t)$. This indicates that the operator $\hat{H}_n(\epsilon_{x_{\vec{n}}} x, \epsilon_{t_{\vec{n}}}t)$ is $\bPC$-symmetric for $(\epsilon_{x_k}, \epsilon_{t_k})=(-1,1)$, $\bT$-symmetric for $(\epsilon_{x_k}, \epsilon_{t_k})=(1,-1)$, and $\bPT$ symmetric for $(\epsilon_{x_k}, \epsilon_{t_k})=(-1,-1),\, k=1,2,...,n$. But for the other types of two family of parameters $(\epsilon_{x_k}, \epsilon_{t_k})$, $\hat{H}_n(\epsilon_{x_{\vec{n}}} x, \epsilon_{t_{\vec{n}}}t)$ must be $\bPC$, $\bT$, or $\bPT$ symmetric.

\begin{table}
\vspace{-0.1in}
\caption{\small $\bPT$ symmetry of the one-component self-induced potential $V_k(x,t)=2\sigma_k q_k^*(\epsilon_{x_k}x,\epsilon_{t_k}t)q_k(x,t)$ for two parameters.\vspace{0.05in}}
 \begin{tabular}{cccccc} \hline\hline \\ [-3.0ex]
 \noindent  Case \quad  &  $(\epsilon_{x_k},\, \epsilon_{t_k})$ \quad & \quad\,\, $\bP$ \,\,\qquad & \quad\,\, $\bPC$ \,\,\quad & \,\,\quad $\bT$ \,\,\quad & \,\,\quad $\bPT$ \qquad  \\ \hline \\ [-3.0ex]
 i &  $(+1, +1)$  &  No &  No & No &  No \\  \hline \\ [-3.0ex]
 ii &  $(-1, +1)$  &  No &  Yes & No &  No \\  \hline \\ [-3.0ex]
 iii &  $(+1, -1)$  &  No &  No & Yes &  No \\  \hline \\ [-3.0ex]
 iv &  $(-1, -1)$  &  No &  No & No &  Yes \\  \hline\hline
\end{tabular}
\label{hnpt}
\end{table}

\section{Multi-linear forms and self-similar reductions}

System (\ref{vnlsp}) admits many types of self-similar reductions. We here give some special reductions:

 {\it Case 1.} If $q_j(\epsilon_{x_j}x,\epsilon_{t_j}t)=\pm q_j(x,t),\, j=1,2,...,n$, then Eq.~(\ref{vnlsp}) reduce to the known local multi-component NLS systems
  whose solutions are known via the Darboux transformation or bilinear method.

 {\it Case 2}. We introduce the rational transformations of Eq.~(\ref{vnlsp})
\bee \label{bi1}
 q_j(x,t)=\frac{g_j(x,t)}{f(x,t)}, \quad f(x,t),\, g_j(x,t)\in\mathbb{C}[x,t]
\ene
which differ from the usual ones~\cite{hirota} such that we substitute transformation (\ref{bi1}) into Eq.~(\ref{vnlsp}) have the combination of  bilinear and multi-linear equations
\bee
\begin{array}{rl} {\rm Bilinear \,\, equation:}& (iD_t+D_x^2-\mu) g_j(x,t)\cdot f(x,t)=0,\quad j=1,2,...,n, \vspace{0.05in} \\
{\rm Multilinear \,\, equation:} & \displaystyle\Pi_{k=1}^nf^{*}(\epsilon_{x_k}x,\epsilon_{t_k}t)(D_x^2-\mu)f(x,t)\cdot f(x,t) \vspace{0.05in}\\
  &\qquad\qquad =-2 f(x,t)\sum_{s=1}^n
  \Big[\sigma_sg_s(x,t)g_j^{*}(\epsilon_{x_j}x,\epsilon_{t_j}t)\Pi_{k=1,k\not=s}^nf^{*}(\epsilon_{x_k}x,\epsilon_{t_k}t)\Big], \end{array}
\label{bi2}\ene
where $\mu\in \mathbb{C}$, $D_t$ and $D_x$ are both Hirota's bilinear operators defined by~\cite{hirota} $D_t^mD_x^n f\cdot g=(\partial_t-\partial_{t'})^m(\partial_x-\partial_{x'})^n[f(x,t)g(x,t)]|_{x=x',t=t'}$. For the case $(\epsilon_x,\epsilon_t)=(1,1)$, we know that
$g_j(\epsilon_{x_j}x,\epsilon_{t_j}t)g_j(x,t)$ are all real-valuable functions such that Eq.(\ref{bi2}) reduces to
\bee\label{bi3}
f^{*}(x, t)(D_x^2-\mu)f(x,t)\cdot f(x,t)=-2 f(x,t)\sum_{j=1}^n\left[\sigma_jg_j(x,t)g_j^{*}(\epsilon_{x_j}x,\epsilon_{t_j}t)\right]
\ene
in which we can assume $f(x,t)\in\mathbb{R}[x,t]$. As a result, Eq.~(\ref{bi3}) becomes a bilinear equation~\cite{hirota}.

{\it Case 3}. Stationary solution reduction. For $\epsilon_{t_j}=1$, Eq.~(\ref{vnlsp}) admits the stationary solutions $q_j(x,t)=\phi_j(x)e^{-i\mu_jt}$ with $\phi_j(x)$ satisfying the system of one-family-parameter $(\epsilon_{x_k})$ ordinary differential equations
\bee
 \mu_j\phi_j(x)=-\phi_{jxx}(x)+2\phi_j(x)\sum_{k=1}^n\sigma_k\phi_k(x)\phi_k^{*}(\epsilon_{x_k}x),\quad \epsilon_{x_k}=\pm 1,\quad k, j=1,2,...,n,
\ene

{\it Case 4}. Symmetry reduction. we use the direct  method~\cite{ck} to consider the  solution of Eq.~(\ref{vnlsp})
\vspace{-0.03in}\bee \label{si1a}
\begin{array}{l}
q_j(x,t)=\frac{p_j(z)}{\sqrt{2t}}e^{i\mu_j\log|t|/2}, \,\,\, z(x,t)=x/\sqrt{2t},
\end{array}
\ene
where $\mu_j\in\mathbb{R}$, $p_j(z)\in \mathbb{C}[z],\, (j=1,2,...,n)$, and $ x,t\in \mathbb{R}$, where $t>0$ is required. Thus the substitution of Eq.~(\ref{si1a}) into Eq.~(\ref{vnlsp}) yields the equation for $p_j(z)$
\vspace{-0.03in} \bee \label{si1b}
  p_{j,zz}-(i+\mu_j)p_j(z)-izp_{j,z}(z)-2p_j(z)\sum_{k=1}^n\sigma_kp_k(z)p_k^{*}(\hat{z}_k)=0,
\ene
where $\hat{z}_k=\epsilon_{x_k} x/\sqrt{2\epsilon_{t_k}t}$,\, k=1,2,...,n. Particularly, i) for $(\epsilon_{x_k}, \epsilon_{t_k})=(1,1),\, k=1,2,...,n$, we have $\hat{z}=z$ and $p_j^{*}(\hat{z})=p_j^{*}(z)$ such that the symmetry  reduction reduces to the known one; ii) for $(\epsilon_{x_k}, \epsilon_{t_k})=(-1,1),\, k=1,2,...,n$, we have $\hat{z}=-z$ and $p_j^{*}(\hat{z})=p_j^{*}(-z)$ such that the symmetry reduction given by Eqs.~(\ref{si1a}) and (\ref{si1b}) becomes the known result~\cite{yanaml15}; iii) for  $(\epsilon_x, \epsilon_t)=(1,-1),\, (-1,-1)$, the symmetry  reductions  given by Eqs.~(\ref{si1a}) and (\ref{si1b}) become the known ones~\cite{yanaml15,yanaml16}.

{\it Case 5}. Separation of variables. we use the separating variables to consider the solution
$q_j(x,t)=p_j(x)e^{\omega_j\sqrt{-\epsilon_{t_j}}t},\, (j=1,2,...,n)$,
where $\omega_j\in \mathbb{R}$,\, $p_j(x)\in \mathbb{C}[x]$, and $ x,t\in \mathbb{R}$. The substitution of this transformation into Eq.~(\ref{vnlsp}) yields the system of ordinary differential equations
 \vspace{-0.05in}\bee\label{si2b}
   p_{j,xx}(x)+i\omega_j \sqrt{-\epsilon_{t_j}} p_j(x)-2p_j(x)\sum_{k=1}^n\sigma_k p_k(x)p_k^{*}(\epsilon_{x_k}x)=0,\,\, j=1,2,...,n
\ene
For $(\epsilon_{x_j}, \epsilon_{t_j})=(1,1)$, we can assume $p_j(x)\in \mathbb{R}[x]$, whereas one of $(\epsilon_{x_j}, \epsilon_{t_j})$ is chosen as $(-1,1),\, (1,-1),\, (-1,-1)$, we may assume $p_j (x)\in \mathbb{C}[x]$ from Eq.~(\ref{si2b}). Other types of symmetry solutions of Eq.~(\ref{vnlsp}) can also be found via the Lie classical and non-classical symmetry methods~\cite{sm}.

%\section{Conclusion and discussion}

In conclusion, we have first introduced a unified two-family-parameter nonlinear evolution hierarchy from a new and simple two-family-parameter symmetry reduction of vector NLS system. The two-family-parameter system contains the local, nonlocal, and novel mixed-local-nonlocal vector NLS equations. Some of the systems with some chosen parameters are shown to admit Lax pairs and infinite number of conservation laws. The whole hierarchy is $\bPT$ symmetric. We also present its similarity reductions such as multi-linear forms, the reduction of separation of variables, and other symmetry solutions. Particularly, the mixed-local-nonlocal system in the $\mQ^{(n)}_{\epsilon_{x_{\vec{n}}},\epsilon_{t_{\vec{n}}}=1}$  and  ${\mathbb Q}^{(n)}_{\epsilon_{x_{\vec{n}}},\epsilon_{\tau_{\vec{n}}}=-1}$ systems present us with new integrable nonlinear evolution equations in the fields of integrable systems. We know that the introduced two families of parameters $(\epsilon_{x_k},\epsilon_{t_k})$ or $(\epsilon_{x_k},\epsilon_{\tau_k}),\,k=1,2,...,n$ play a central role in the study of both new integrable systems and $\bPT$ symmetry. In fact, the idea for two-family-parameter used in this paper can also be extended to other nonlinear evolution equations (e.g., vector KdV equations, vector mKdV equations, vector sine-Gordon equations, the higher-order AKNS hierarchy, the higher-dimensional $(x,y,z)$~\cite{hd},  discrete systems~\cite{vnlsb,amd}, etc.) such that the corresponding local, nonlocal, and mixed local-nonlocal systems can be found. For example, we have the fourth-order AKNS equation
\bee\begin{array}{l}
 iq_t+\alpha\big[\frac{1}{2}q_{xx}+q^2(x,t)q^*(\epsilon_xx, \epsilon_tt)\big]+ \gamma \big[q_{xxxx}+8q(x,t)q^*(\epsilon_xx, \epsilon_tt)q_{xx}(x,t) +6q(x,t)^3q^{*2}(\epsilon_xx, \epsilon_tt)q_{xx}(x,t)\v\\
 \quad\,+4q_x(x,t)q_x^*(\epsilon_xx, \epsilon_tt)q(x,t)+6q^*(\epsilon_xx, \epsilon_tt)
 q_{x}^2(x,t)+2q^2(x,t)q_{xx}^{*}(\epsilon_xx, \epsilon_tt)\big]=0,
\end{array}
\ene
and other the even-order AKNS equations. Similarly, a unified new two-family-parameter discrete local-nonlocal vector hierarchy can be given as
\bee \begin{array}{l} i{\bf P}_{n,t}(t)=-{\bf P}_{m+1}(t)+2{\bf P}_{m}(t)-{\bf P}_{m-1}(t)
+\sum_{j=1,2}{\bf P}_{m+2-j}(t){\bf P}_{\epsilon_{x_{\vec{n}}} m}^{\dag}(\epsilon_{t_{\vec{n}}}t){\bf P}_{m+1-j}(t) \end{array} \ene
 with $m\in \mathbb{Z}$,\, ${\bf P}_m(t)=(p_{1,m}(t),\, p_{2,m}(t), ...,
p_{n,m}(t))^T$ and ${\bf P}_{\epsilon_{x_{\vec{n}}} m}(\epsilon_{t_{\vec{n}}}t)=(p_{1, \epsilon_{x_1}m}(\epsilon_{t_1}t),\, p_{2, \epsilon_{x_2}m}(\epsilon_{t_2}t), ..., p_{n, \epsilon_{x_n}m}(\epsilon_{t_n}t))^T$,  ${\bf P}_{\epsilon_{x_{\vec{n}}} m}^{\dag}(\epsilon_{t_{\vec{n}}}t)$ stands for the transpose conjugate of ${\bf P}_{\epsilon_{x_{\vec{n}}} m}(\epsilon_{t_{\vec{n}}}t)$, which is regarded as a kind of discretization of Eq.~(\ref{vnlsp}).
The wave structures of these obtained new systems  will be explored in another literature. 

Finally, it should be pointed out that we present some novel Lax-integrable mixed-local-nonlocal vector NLS equations using above-mentioned way. But we present an open problem whether the {\it single mixed-local-nonlocal NLS equation}
\bee
iq_{t}=-q_{xx}+\sigma_1|q|^2q+\sigma_2q^2q^*(\epsilon_{x_1}x,\epsilon_{t_1}t), \quad 
\epsilon_{x_1}=e^{i\theta_{x_1}},\epsilon_{t_1}=e^{i\theta_{t_1}},\, \theta_{x_1}, \theta_{t_1}\in[0, 2\pi),\,
\ene
is Lax-integrable, where $q=q(x,t)$, $\sigma_1$ is the nonlinear coefficient of local-local interaction, and $\sigma_2$ is the nonlinear coefficient of local-nonlocal interaction. In  other word, whether does there exist a Lax-integrable {\it single nonlinear model} including
both local and nonlocal nonlinearities (i.e., both local-local and local-nonlocal/nonlocal-nonlocal interactions) ?  

\vspace{0.1in} \noindent {\bf Acknowledgments}

This work was partially supported by the NSFC under Grant Nos. 11571346 and 11731014.

%%%%%%%%%%%%%%%%%%%%%%%%%%%%%%%%%%%%%%%%%%%%%%%%%%%%%%%%%%%%%%

%\begin{thebibliography}{99}

%\vspace{-0.25in}

\end{document}